# Bursts of Active Transport in Living Cells

Bo Wang,[a] James Kuo,[b] and Steve Granick[a,b,c,d,1]

*Departments of Materials Science,[a] Chemical and Biomolecular Engineering,[b]*

*Chemistry,[c] and Physics[d]*

*University of Illinois, Urbana, IL 61801 USA*

[1] *To whom correspondence should be addressed. E-mail: sgranick@uiuc.edu*

## Abstract

We scrutinize the temporally-resolved speed of active cargo transport in living cells, and show intermittent bursting motions.   These nonlinear fluctuations follow a scaling law over several decades of time and space, the statistical regularities displaying a time-averaged shape that we interpret to reflect stress buildup followed by rapid release.  The power law of scaling is the same as seen in driven jammed colloids, granular, and magnetic systems.  The implied regulation of active transport with environmental obstruction extends the classical notion of "molecular crowding".

One of the most fundamental differences between the inanimate and living world is the prominence of active transport in living systems, driven by the proteins known as molecular motors [1,2]. This molecular machinery underpins essential functions, among them locomotion, cell cycle, signaling and metabolism. Much is known about the molecular mechanisms of elementary steps taken by the motors [3-6], but not enough about how these steps coordinate in the context of crowded cellular environment. Here, analyzing a uniquely large dataset of individual trajectories of cargo transport in living cells, we reveal statistical commonality with driven motion in jammed nonliving systems and a conceptual connection to delocalization in glassy or jammed dynamics [7-13].

What is known already about active transport is that, when a single motor protein (say, kinesin or dynein) drags a vesicular cargo, such as endosome or lysosome, along a microtubule, the motor is subject to an opposing load force arising from viscous drag (Fig. 1a, top panel). Extensive exploration *in vitro* shows that the speed decreases approximately linearly with respect to increase of the load force and that multiple motors can bind to the cargo simultaneously with load-dependent dissociation rates [5,6]. Thus, mobility presumably is determined by the winning side of two opposing forces (Fig. 1a, middle panel) [5,6]. Here we consider the complexity of cargo that must be dragged through the crowded, heterogeneous cytoplasm of a living cell (Fig. 1a, bottom panel). If the environment just presents viscous drag, in existing theoretical models, the cargo speed is expected to take discrete values according to the number of molecular motors, with some noise around these fixed points (Fig. 1b, middle panel) [5]. Alternatively (Fig. 1b, bottom panel), as cargo is pulled through a crowded environment, we hypothesize that speed would decrease as stress builds up, the environment pushing against cargo motion; then, bursting free of this transient confinement, cargo would speed up. Thus, while it is known that on short time and length scales, the speed is dictated by the molecular kinetics [5,6] (~1 ms, ~10 nm), slow environmental structural rearrangement might contribute significantly on larger scales. Therefore, the slow *fluctuation* of the speed is expected to be instructive, but earlier studies in living cells focused instead on the *average* speed.

To follow active transport in living cells at the single-cargo level, we used fluorescence imaging. Living mammalian cells were imaged under physiological conditions in a homebuilt microscope using highly laminated illumination (HILO) [14] to improve the signal-to-noise ratio and minimize possible error introduced by 2D projection. To fluorescently label specific populations of endosome cargo, either 0.15 µg/mL biotinylated epidermal growth factor (EGF) complexed to Alexa-555 streptavidin or 10 µg/mL DiI conjugated low-density lipoprotein (LDL) (Invitrogen) was applied to cells. To fluorescently label lysosome cargo, cells were transiently transfected with lysosomal-associated membrane protein 1 (LAMP1) fused to tagRFP (Invitrogen). Fluorescence images were taken on the basal side of cells at frame rate of 50 ms, this time being selected as longer than the motor stepping time (1-10 ms) to remove molecular noise. The center positions of the cargos were each located with a resolution of <5 nm and strung together by an automated program to form trajectories [15].

The active transport portions of trajectories were identified by wavelet analysis [16] and the speed *along* the (sometimes-curved) active transport path was calculated over 50 ms time intervals. To confirm self-consistency of the data, we inspected displacements *perpendicular* to the track, and found them to be Gaussian in distribution with variance of 20 nm; this reasonable number coincides with the sum of microtubule radius and length of motor stalk, indicating perpendicular displacements are mainly contributed from transverse swinging of cargos on our time scale. Furthermore, when 5 µg/mL nocodazole was supplemented in the medium to disassemble microtubules, no active transport was observed.  This confirmed that the active transport reported below was along microtubules.

Position, velocity, and acceleration are illustrated by raw data in Fig. 2.  We tracked a large number of cells, >50 cells for each experimental condition, and the patterns of data in this figure were verified for >10,000 runs of uninterrupted motion in one direction under each experimental condition.  Panel a shows a typical run with color denoting elapsed time, one notices that motion is sometimes slow, sometimes fast.  Panel b shows two examples of temporally-resolved speed versus time;  one notices asymmetric bursting dynamics, different from ordinary noise around a mean value.  Panel c plots speed against acceleration over a large range of variables (left plot) and with a magnified view of the same data (right plot).  The spiral patterns show that speed fluctuations have a nonlinear dynamic feature with memory as long as one second, a time that coincides with the known typical mechanical relaxation time of cells [13].  Limit cycles are not observed, so this motion is not periodic, but the circuits are strikingly parallel and tilted.  In other words, while fluctuations differ in both amplitude and time, their shapes are similar and they are regularly asymmetrical with time.  Speed fluctuations do not appear to cluster around distinct values, nor do they present distinct peaks of multiple Gaussian distributions, features that would be expected if they reflect motor association-dissociation events [5,17].

Next, we considered correlations.  Ensemble-averaged speed-speed autocorrelation functions of active cargo transport are plotted in Fig. 3 against time.  The negative correlation peak is pronounced at ~0.3 s, yet negative over a significantly broader time, confirming the long memory already noted.  Physically, we interpret this to indicate a broad distribution of the time scales of fluctuations, a ubiquitous feature of dynamic heterogeneity.  This is in contrast to a sharp peak. A sharp peak (not observed) would correspond to a well-defined time scale, which would be anticipated if the observed nonlinearity were induced by a specific molecular biochemical cycle.

For further quantification, we analyzed the duration and travel distance of each fluctuation cycle. Given a time series of speed $\upsilon(t)$ for a single run, we imposed a reference $\tilde{\upsilon}$ defined as 60% of the most likely speed in this run, and defined "bursts" as periods of time when speed exceeded this reference. Each burst has a duration $T$ measured as the time between two successive intersections of $\upsilon(t)$ with $\tilde{\upsilon}$. Each burst also has a distance traveled within this run, the burst length $L$. Fig. 4a (top) shows that when the rather noisy spectrum of raw data is averaged, the striking phenomenological power law is observed over more than 2 decades of $T$, $L \sim T^{\frac{3}{2}}$. It is not known whether the slight bending down at longest $T$ might reflect limited statistics. This power of 3/2, independent of how the reference speed $\tilde{\upsilon}$ is defined, matches phenomenologically that expected from crackling noise in critical inanimate systems [7,18].

Importantly, modulating the cytoskeleton with latrunculin A (LatA) or cytochalasin D (CytoD), both of which dissemble actin filaments, or Y-27632, which perturbs cytoskeleton by inhibiting upstream regulator Rho-associated protein kinase, switches to 2 the scaling power of bursts less than $L = 100$ nm and $T = 0.3$ s (Fig. 4a, bottom panel). It is probably significant that $L = 100$ nm coincides with the known intracellular mesh size of actin networks. Above this length scale the impedance from cytoskeleton may average out and mix with contributions from other cytoplasmic components; it is reasonable to observe differences only for $L$ smaller than this. Consistently, after actin disruption, the first minimum in speed-speed autocorrelation curves becomes shallower, indicating a shift towards longer times (Fig. 3b). Likewise, the scaling length $L_0$ shifts to longer distances, as it provides a typical length scale of the bursts under that condition. This is plotted (Fig. 4b) for 11 types of experiments in which system-specific details such as cargo size, number of motors bound, and molecular regulation were all varied. Taken together, these observations indicate that disruption of actin network loosens the constraint on cargo transport. Regardless, after time $t$ has been rescaled by burst-specific duration $T$ and speed by $T^{\frac{1}{2}}$ (Fig. 4c), all bursts fall on the same master curve.

The asymmetry in this time-averaged data is consistent with the picture that as cargo moves, dragged by molecular motors, at first stress builds up such that motion slows, then it releases and motion speeds up (Fig. 4c left panel). The initial acceleration during which speed increases linearly with time (the magnified view in Fig. 4c right panel) may reflect release of elastic stress. The power-law with slope *3/2* is consistent with the simple heuristic argument that these data reflect the superposition of ballistic motion and random Brownian noise [19]. Taken together, although it is true that at the molecular level this endogenous transport surely involves complex regulatory pathways and genes, and it is also true that the exact shape of stress relaxation should depend on the detailed relaxation mechanism of the environment, the striking power laws demonstrated here suggest scale-free simplicity, when coarse-grained by the normalizations we have made. This striking invariance highlights the redundancy in regulation of cellular active transport, which in turn relates to universality in cell mechanics that has been proposed by others [12,13]. All this was insensitive to changing the molecular details.

In summary, temporally-resolved speed of active transport in living cells shows striking statistical regularities. The universality of the scale-free power law scaling observed here is in quantitative agreement with those reported in inanimate systems (jammed colloids and granular media, and magnetic Barkhausen noise [7,18], suggesting a common origin in pushing through a crowded environment in weak force regime.

We thank Jianjun Cheng for allowing us to use his cell culture facility, Dongwan Yoo for providing the Marc-145 monkey kidney cell line, and Stephen M. Anthony for the single particle tracking program. This work was supported by the U.S. Department of Energy, Division of Materials Science, under Award DEFG02-02ER46019.

# Figure legends

**Figure 1.** The main idea of this experiment. (a) Alternative hypothetical scenarios for the rate-limiting step of active vesicular transport along microtubules. Top: the vesicle is dragged by a single motor with speed, $v$, determined by the load force, $F$. Middle: the vesicle is dragged by multiple motors whose number fluctuates. Bottom: the environment blocks motion transiently. Colors denote cytoskeleton (yellow) and cytoplasm components (purple). (b) Hypothetical speed fluctuations under these scenarios. Top: the speed fluctuates around a mean with Gaussian noise. Middle: the speed fluctuates between distinct levels corresponding to different numbers of working motors, with Gaussian noise over shorter times. Bottom: fluctuations resemble bursts as cargo squeezes through an obstructing environment.

**Figure 2.** Speed fluctuations of endosomes transported along microtubules. (a) A representative uninterrupted "run" lasting 5 s of EGF-containing endosomes in HeLa cells, circles representing the raw position data obtained from tracking, the color denoting elapsed time. Positions with constant time step (50 ms) along the contour of the track are shown on the bottom horizontal line. The arrows identify the demarcation between slow periods separated by fast bursts. (b) Two representative runs of EGF-containing endosomes in HeLa cells with temporally-resolved speed, evaluated over 50 ms intervals, plotted versus time. The fluctuation is one order of magnitude larger than the measurement uncertainty, 0.2 μm/s. (c) The same two time series are plotted as instantaneous speed versus acceleration (left) with a magnified view near the origin (right).

**Figure 3.** The ensemble-averaged speed-speed autocorrelation function $R(\Delta t)$ of active transport. (a) The shape of $R(\Delta t)$ vs. $\Delta t$ is invariant to experimental conditions. (b) The effect of cytoskeleton perturbations. The symbol representations are summarized in Fig. 4. The solid line depicts the average curve for cytoskeleton-intact samples. The dashed line is a single exponential fit. To compute the autocorrelation consistently, we cropped runs longer than 50 frames into pieces of length 50 frames or shorter.

**Figure 4.** Scaling of speed fluctuations for the active transport of endosomes and lysosomes. (a) Top: Log-log plot of burst length ($L$) versus their duration ($T$); these data refer to EGF-labeled endosomes in HeLa cells. Averaged from 26,496 bursts, the crosses (raw data shown as grey circles), have the slope indicated in the figure. Bottom: Collapse of such data after modulating the cytoskeleton when comparing $L/L_o$, where $L_o$ differs between systems. The solid lines have the slopes indicated in the figure, highlighting a change of the power law when $L < 100$ nm and $T < 0.3$ s. (b) The length $L_o$ inferred from 11 systems under different conditions. The dashed lines indicate systematic shift of $L_o$ to longer length upon actin disruption. The experiments are EGF containing endosomes in, 1, HeLa cells, 2, Marc145 cells, and, 3, DU145 cells; 4, LAMP1 labeled lysosomes in HeLa cells; LDL containing endosomes in, 5, HeLa cells, and, 6, DU145 cells; EGF containing endosomes in HeLa cells, 7, with microtubule-associated protein 4 (MAP4) overexpression (MAP4 is known to act as a road blocker on microtubules), 8, treated with 5 µM trichostatin A (TSA inhibits histone deacetylase, inducing microtubule acetylation that enhances the processivity of the motors); 9, treated with 50 µM Y27632; 10, with 0.08 µg/mL LatA; 11, with 0.75 µg/mL CytoD. For LatA and CytoD experiments, the concentrations were chosen such that blebs formed at cell peripheries but cells retained their shapes and no complete rounding was observed within 3 h.  Symbol representations are consistent with previous figures. (c) The average burst shape, obtained by normalizing all bursts longer than 0.3 s to the same duration ($T$) and area (left). Compared to the solid line, a semicircle, the asymmetry of the data is evident. The tail for small fractional period, $t/T$, is plotted on log-log scales (right). The dashed line through the data has slope of 1 indicating an accelerating stage.  In comparison, the tail of the semicircle (solid line) has slope close to 1/2.  Importantly, the data under all conditions collapse.

Figure 1

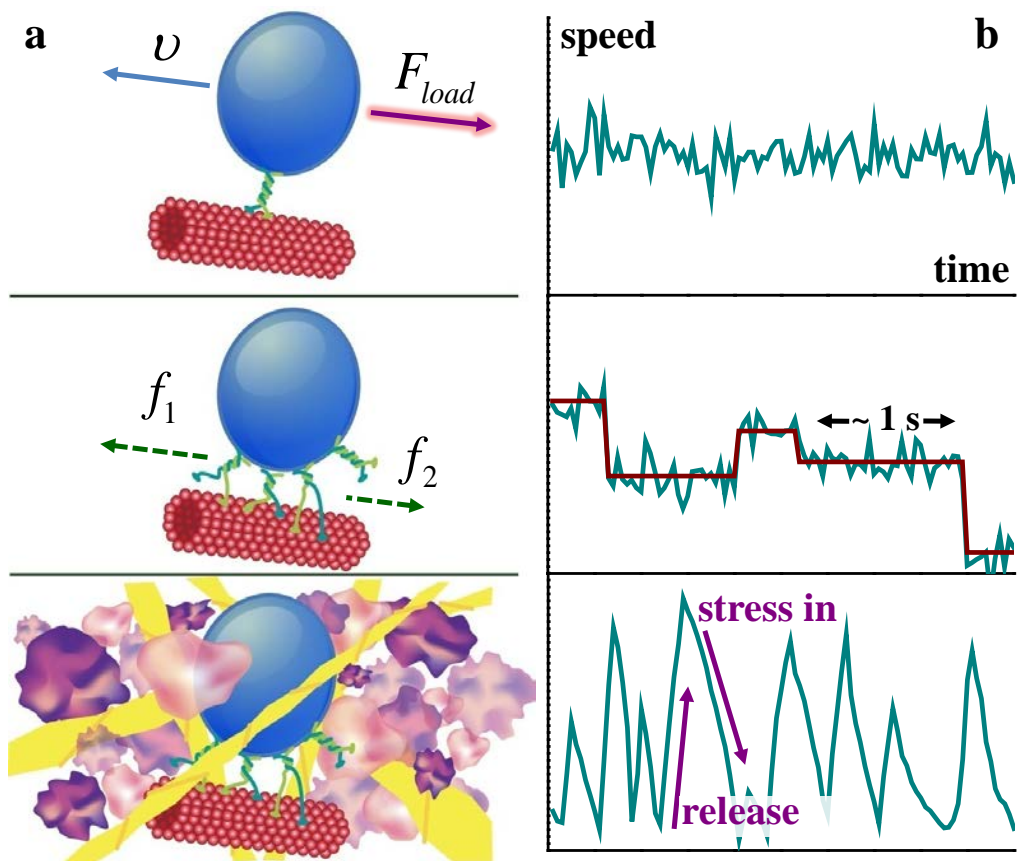

Figure 2

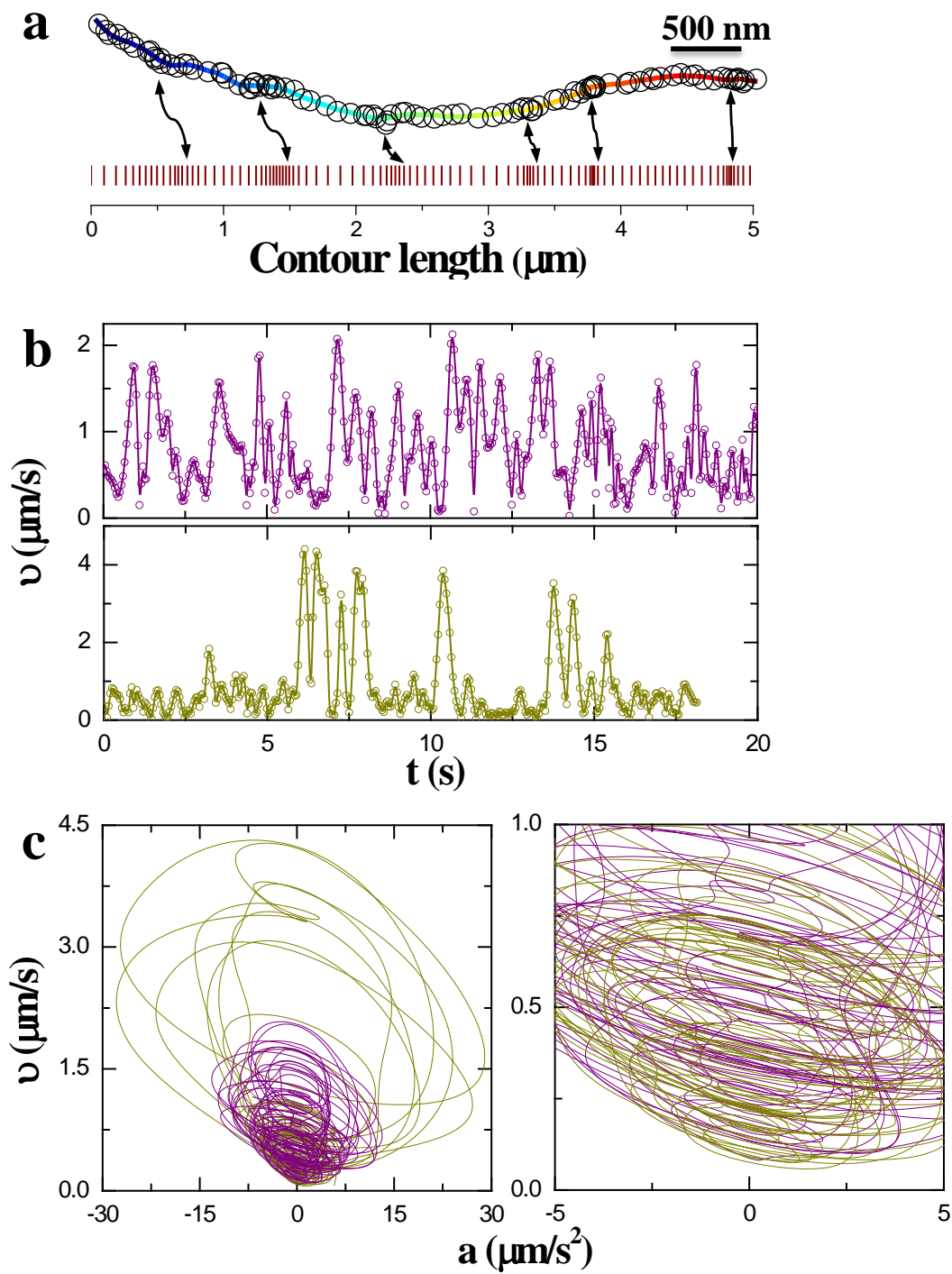

Figure 3

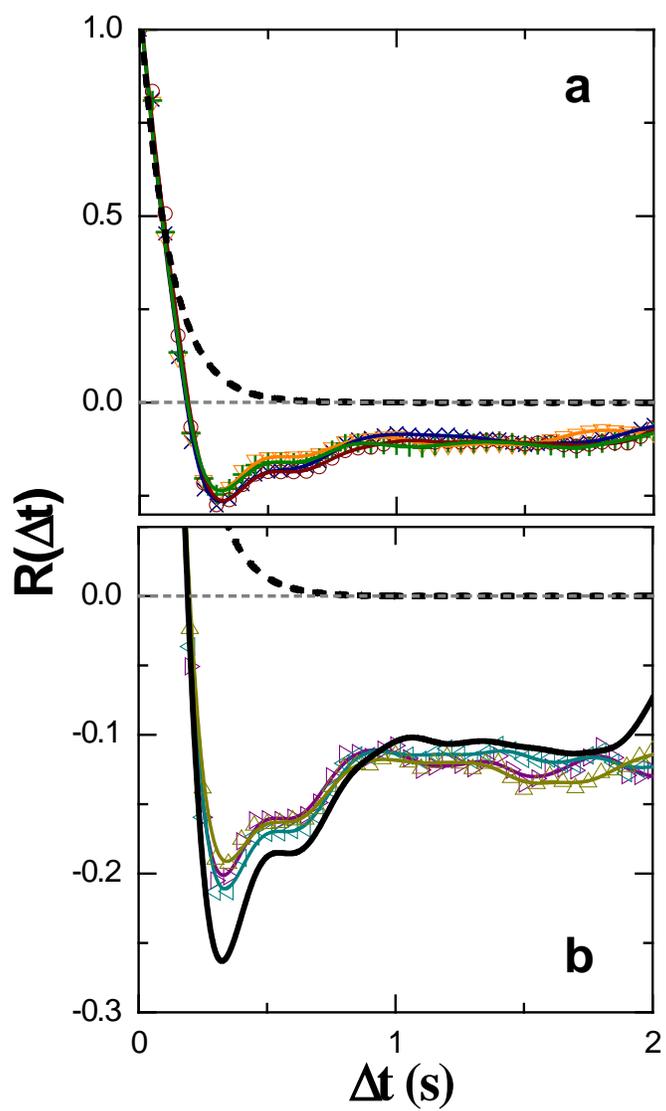

Figure 4

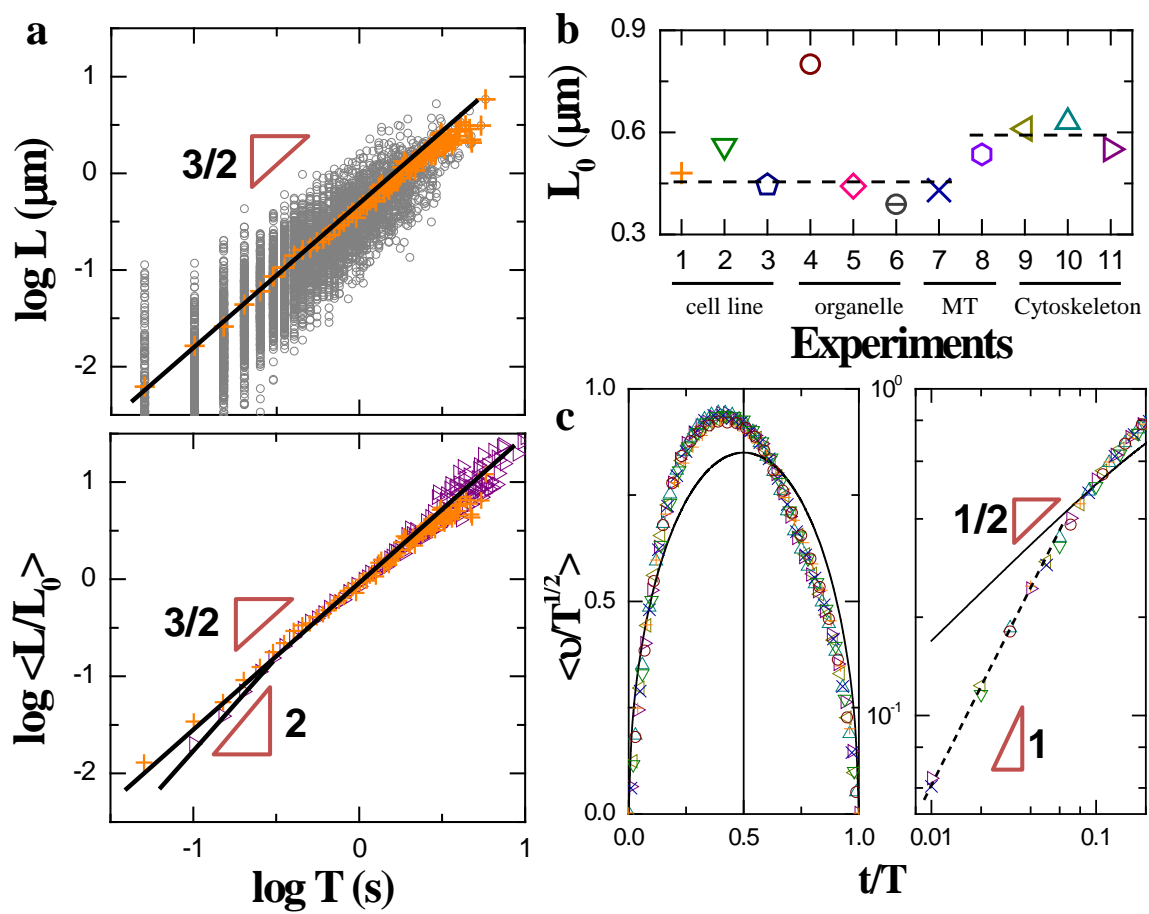